\documentclass[conference]{IEEEtran}
\IEEEoverridecommandlockouts
\usepackage{cite}
\usepackage{amsmath,amssymb,amsfonts}
\usepackage{algorithmic}
\usepackage{graphicx}
\usepackage{textcomp}
\usepackage{xcolor}
\usepackage{listings}
\usepackage{xcolor}
\usepackage{censor}
\usepackage{array}
\usepackage{listings}
\usepackage{xcolor}

\newcolumntype{C}[1]{>{\raggedright\arraybackslash}p{#1}}

\usepackage{booktabs} % For better table formatting
\def\BibTeX{{\rm B\kern-.05em{\sc i\kern-.025em b}\kern-.08em
    T\kern-.1667em\lower.7ex\hbox{E}\kern-.125emX}}
\begin{document}

\title{WIP: Leveraging LLMs for Enforcing Design Principles in Student Code: Analysis of Prompting Strategies and RAG}

\author{
\IEEEauthorblockN{Dhruv Kolhatkar}
\IEEEauthorblockA{\textit{Department of Computer Science} \\
\textit{North Carolina State University}\\
Raleigh, NC, USA\\
dukolhat@ncsu.edu}
\and
\IEEEauthorblockN{Soubhagya Akkena}
\IEEEauthorblockA{\textit{Department of Computer Science} \\
\textit{North Carolina State University}\\
Raleigh, NC, USA\\
sakkena@ncsu.edu}
\and
\IEEEauthorblockN{Edward F. Gehringer}
\IEEEauthorblockA{\textit{Department of Computer Science} \\
\textit{North Carolina State University}\\
Raleigh, NC, USA\\
efg@ncsu.edu}
}

\maketitle

\begin{abstract}
This work-in-progress research-to-practice paper explores the integration of Large Language Models (LLMs) into the code-review process for open-source software projects developed in computer science and software engineering courses. The focus is on developing an automated feedback tool that evaluates student code for adherence to key object-oriented design principles, addressing the need for more effective and scalable methods to teach software design best practices. The innovative practice involves leveraging LLMs and Retrieval-Augmented Generation (RAG) to create an automated feedback system that assesses student code for principles like SOLID, DRY, and design patterns. It analyzes the effectiveness of various prompting strategies and the RAG integration. Preliminary findings show promising improvements in code quality. Future work will aim to improve model accuracy and expand support for additional design principles.
\end{abstract}


\section{Introduction}
Software design principles are foundational to creating maintainable, scalable, and efficient systems \cite{Martin2017}. Despite their critical importance, effectively teaching and consistently enforcing these principles in an educational setting remains a persistent challenge. Students often struggle to translate abstract design concepts into practical code, leading to systems that are difficult to understand, extend, or debug. The traditional methods of providing feedback on design adherence, such as manual code reviews, are often time-consuming, resource-intensive, and inherently limited in their scalability and consistency \cite{Piech2015, Keuning2019}. This bottleneck significantly impacts the quality of software engineering education, as students may not receive the timely and targeted feedback necessary to internalize best practices.

Recent advancements in Large Language Models (LLMs) open new and exciting avenues for automated code review, offering the potential to provide contextually relevant and timely feedback to students at scale. This work-in-progress research-to-practice paper investigates the potential of LLMs, combined with Retrieval-Augmented Generation (RAG) \cite{Lewis2020}, to enhance feedback on student code in software engineering courses. By leveraging RAG, LLMs gain access to a curated knowledge base of established design principles, significantly improving their ability to assess code in line with widely accepted best practices. This integration not only promises to make the feedback process more efficient but also aims to enrich the learning experience by providing students with immediate, principle-grounded insights into their code's design quality. Our research explores various prompting strategies to optimize LLM performance and analyzes the effectiveness of RAG in enhancing the specificity and educational value of the generated feedback, ultimately addressing a long-standing need for more effective and scalable methods to teach software design best practices in computer science and software engineering curricula.

\section{Related Work}
The integration of artificial intelligence into educational tools is not new\cite{VanLehn2006, Anderson1985}; intelligent tutoring systems and AI-assisted learning platforms have long aimed to support students in mastering programming concepts. However, limited work has explored the use of cutting-edge LLMs for teaching complex object-oriented design principles\cite{Larman2004}. This research extends prior efforts in automated code review and educational tooling by investigating how prompting strategies\cite{Brown2020,Reynolds2021,Wei2022} and RAG can be leveraged to teach software design more effectively.

\section{Methodology}
We evaluated GPT, Claude, and DeepSeek using six custom prompts designed to assess different facets of software design quality. These ranged from class-level analysis to architectural critique, incorporating principles like SOLID, DRY, KISS, and targeting readability, maintainability, and design flaws.
\begin{itemize}
    \item \textit{Prompt A:} Analyzes class relationships and suggests improvements in code design and coupling, helping identify areas where responsibilities are unclear or where dependencies can be decoupled to improve modularity\cite{Parnas1972}.

    \item \textit{Prompt B:} Evaluates code using a comprehensive set of design principles and best practices:
    \begin{itemize}
        \item \textit{SOLID}: Ensures classes are well-structured, promoting single responsibility, open/closed design, and minimal interdependencies.
        \item \textit{DRY (Don't Repeat Yourself)}: Identifies code duplication and suggests abstraction to improve maintainability.
        \item \textit{KISS (Keep It Simple, Stupid)}: Encourages simplicity by flagging unnecessary complexity or overengineering.
        \item \textit{YAGNI (You Aren’t Gonna Need It)}: Discourages premature optimizations or unused features that increase codebase bloat.
        \item \textit{PoLA (Principle of Least Astonishment)}: Promotes intuitive design and naming, ensuring code behaves in expected ways.
        \item \textit{Law of Demeter}: Encourages low coupling by restricting how objects interact with nested dependencies (“only talk to your immediate friends”).
        \item \textit{SoC (Separation of Concerns)}: Supports modularization by separating logic into cohesive units (e.g., separating controllers, models, and services).
        \item \textit{Design Patterns}: Assesses adherence to commonly accepted architectural solutions (e.g., Factory, Strategy, MVC) for recurring design challenges\cite{Gamma1994}.
        \item \textit{Readability}: Evaluates whether the code is clear and understandable, with logical naming, consistent formatting, and appropriate documentation.
    \end{itemize}

    \item \textit{Prompt C:} Compares code against broader industry best practices, focusing on readability, maintainability, adherence to design principles, detection of code smells, and support for modular development. It provides high-level architectural insights and flags anti-patterns or structural flaws.

    \item \textit{Prompt D:} Critiques code with a focus on readability issues, including excessive nesting, poor naming, and complexity.
    \item \textit{Prompt E:} Analyzes potential design flaws, focusing on code smells and bad design patterns.
    \item \textit{Prompt F:} Provides a holistic architectural analysis, focusing on class design, maintainability, and overall code quality.
\end{itemize}
Unfortunately, the complete prompts are too long to fit in the paper, but the authors will provide them upon request.

\section{Comparative Model Performance}

We evaluated the performance of three LLMs---GPT-4, Claude, and DeepSeek---across six structured prompts, each targeting different dimensions of software quality, including design principles, architecture, readability, and maintainability. The results are presented below, followed by a detailed comparative analysis.

\subsection{Quantitative Results}
The following tables summarize the number of issues each model identified, the total unique issues per prompt, and the corresponding percentage of coverage.  For example, for Prompt A, a total of 14 issues were identified by the various models.  No model found all 14; DeepSeek found the most (9), which amounted to 64.3\% of the total issues.  
\vspace{1em}
\begin{table}[htbp!]
\centering
\caption{Prompt A: General Code Analysis Prompt Results}
\begin{tabular}{lccc}
\toprule
Model & \# Issues Identified / Total Unique Issues & \% Total Issues  \\
\midrule
GPT & 3 / 14 & 21.4\% \\
Claude & 8 / 14 & 57.1\% \\
DeepSeek & 9 / 14 & 64.3\% \\
\bottomrule
\end{tabular}
\label{tab_prompt_a}
\end{table}

\begin{table}[htbp!]
\centering
\caption{Prompt B: Principle-Focused Prompt, Results}
\begin{tabular}{lccc}
\toprule
Model & \# Issues Identified / Total Unique Issues & \% Total Issues  \\
\midrule
GPT & 5 / 17 & 29.4\% \\
Claude & 7 / 17 & 41.2\% \\
DeepSeek & 6 / 17 & 35.3\% \\
\bottomrule
\end{tabular}
\label{tab_prompt_b}
\end{table}

\begin{table}[htbp]
\centering
\caption{Prompt C: Industry Best Practices Prompt Results}
\begin{tabular}{lccc}
\toprule
Model & \# Issues Identified / Total Unique Issues & \% Total Issues  \\
\midrule
GPT & 8 / 31 & 25.8\% \\
Claude & 16 / 31 & 51.6\% \\
DeepSeek & 12 / 31 & 38.7\% \\
\bottomrule
\end{tabular}
\label{tab_prompt_c}
\end{table}

\begin{table}[htbp!]
\centering
\caption{Prompt D: Readability-Focused Prompt Results}
\begin{tabular}{lccc}
\toprule
Model & \# Issues Identified / Total Unique Issues & \% Total Issues  \\
\midrule
GPT & 8 / 28 & 28.6\% \\
Claude & 16 / 28 & 57.1\% \\
DeepSeek & 12 / 28 & 42.9\% \\
\bottomrule
\end{tabular}
\label{tab_prompt_d}
\end{table}

\begin{table}[htbp!]
\centering
\caption{Prompt E: Code Smells and Design Flaws Prompt Results}
\begin{tabular}{lccc}
\toprule
Model & \# Issues Identified / Total Unique Issues & \% Total Issues  \\
\midrule
GPT & 5 / 15 & 33.3\% \\
Claude & 6 / 15 & 40.0\% \\
DeepSeek & 5 / 15 & 33.3\% \\
\bottomrule
\end{tabular}
\label{tab_prompt_e}
\end{table}

\begin{table}[htbp!]
\centering
\caption{Prompt F: Holistic Architectural Analysis Prompt Results}
\begin{tabular}{lccc}
\toprule
Model & \# Issues Identified / Total Unique Issues & \% Total Issues  \\
\midrule
GPT & 6 / 15 & 40.0\% \\
Claude & 5 / 15 & 33.3\% \\
DeepSeek & 6 / 15 & 40.0\% \\
\bottomrule
\end{tabular}
\label{tab_prompt_f}
\end{table}

\subsection{Prompt-Wise Model Analysis}

\textit{Prompt A, Code structure and coupling:} 
DeepSeek provided the most thorough analysis (9 relevant issues), identifying empty classes, improper ORM practices, and code organization flaws. Claude closely followed with a similarly comprehensive review (8 issues). GPT, while briefer, focused on key relational design improvements.

\textit{Prompt B, SOLID, DRY, KISS Principles:}  
Claude led with detailed critiques mapped to design principles like SRP and PoLA. GPT accurately flagged excessive number of method calls and poor modularity. DeepSeek also performed well, but its responses were slightly less detailed.

\textit{Prompt C, Industry best practices comparison:}  
Claude stood out with an exhaustive audit (16 issues), covering magic numbers, documentation gaps, and inheritance misuse. GPT gave a clear, practical critique, and DeepSeek was concise yet relevant, particularly in pointing out readability issues and refactor opportunities.

\textit{Prompt D, Readability focused critique:}  
Claude provided the deepest readability audit with detailed issues across naming, nesting, and code comments. GPT and DeepSeek both offered actionable feedback with practical refactoring suggestions.\cite{Fowler2018}

\textit{Prompt E, Design Flaws and Code Smells:}  
GPT clearly identified design flaws like god classes and unnecessary inheritance, offering code-level fixes. Claude added value through Rails-specific suggestions, while DeepSeek effectively highlighted repetition and magic values with a modular lens.

\textit{Prompt F, Holistic architectural analysis:}  
GPT's systemic critique stood out, addressing both architectural and class-level problems. Claude's ActiveRecord-focused suggestions (e.g., enum use, proper ORM relationships) enhanced framework-specific depth. DeepSeek remained consistent, emphasizing maintainability and modularity.

\subsection{Summary of apparent LLM strengths}

\begin{itemize} \item \textit{GPT-4:} GPT-4 excels in architectural analysis and evaluation based on software design principles. It is particularly effective at identifying structural issues in the code and proposing practical refactorings that align with best practices.

\item \textit{Claude:} Claude provides the most comprehensive and detailed analysis among the models. Its strong awareness of the Rails framework makes it exceptionally well-suited for in-depth inspections and nuanced critiques of codebases.

\item \textit{DeepSeek:} DeepSeek offers a balanced and consistent performance. It delivers concise yet insightful feedback, particularly effective in detecting DRY violations and offering clear suggestions to improve readability and modularity. \end{itemize}

\noindent In conclusion, Claude leads in detailed diagnostics, GPT-4 provides sharp architectural insights, and DeepSeek delivers practical, well-scoped analysis.

\section{Effective prompting patterns}
Our evaluation revealed that the structure and phrasing of prompts had a notable impact on LLM performance. Prompts following specific patterns consistently led to more relevant, accurate, and educational responses. The following patterns were particularly effective:

\begin{itemize}

\item \textit{Explicit principle enumeration:}
Listing design principles (e.g., SOLID, DRY) guided the model’s focus.

\textit{Example prompt:}
{\fontfamily{lmr}\selectfont Review this code for violations of SOLID, DRY, and KISS principles.}

\item \textit{Structured format requirements:}
Asking for responses in a specific structure improved clarity.

\textit{Example prompt:}
{\fontfamily{lmr}\selectfont For each issue: (1) Violated Principle, (2) Problem Snippet, (3) Suggested Fix.}

\textit{Sample output:}
Violated Principle: SRP
Problem: Logging inside Controller
Fix: Move logging to LoggerService

\item \textit{Balance between specific and holistic analysis:}
Combining method-level critique with system-wide recommendations produced more comprehensive feedback.

\textit{Example prompt:}
{\fontfamily{lmr}\selectfont Suggest improvements for this method and also comment on the overall class design.}

\item \textit{Framing in an educational context:}
Framing the prompt as a teaching scenario led to more constructive explanations.

\textit{Example prompt:}
{\fontfamily{lmr}\selectfont Explain this issue to a junior developer and suggest how to fix it.}

\textit{Sample output:}
This method is doing too many things (violating SRP). It would be clearer if data fetching and validation were in separate methods.

\end{itemize}

In future applications of LLMs to code review, incorporating these prompting strategies can help maximize both output quality and practical utility.

% \section{Retrieval-Augmented Generation (RAG) Analysis}
% \subsection{RAG Implementation}
% RAG was implemented by providing core design principles as context, enhancing LLM prompts.
% \subsection{RAG Results Analysis}
% RAG improved context-awareness, provided specific references to design principles, and enhanced refactoring suggestions.
\section{Retrieval-Augmented Generation (RAG) Analysis}

\begin{table*}[ht]
\centering
\caption{Refactoring Examples: Comparison Across LLMs}
\renewcommand{\arraystretch}{1.3}
\scriptsize
\begin{tabular}{|C{0.7cm}|C{4.1cm}|C{4.1cm}|C{3.7cm}|C{3.8cm}|}
\hline
\textbf{Number} & \textbf{Original Code} & \textbf{RAG (GPT-4)} & \textbf{Claude} & \textbf{DeepSeek} \\
\hline
1 & 
\begin{lstlisting}
def check_topic_due_date_val
(assignment_due_dates, topic_id, deadline_type_id = 1, review_round = 1)
  ...
end
\end{lstlisting} &
\begin{lstlisting}
def check_topic_due_date_val
(assignment_due_dates:, topic_id:, deadline_type_id: 1, review_round: 1)
  ...
end
\end{lstlisting} &
\begin{lstlisting}
class TopicDeadlineService
  def formatted_due_date(topic_id)
    ...
  end
end
\end{lstlisting} &
\begin{lstlisting}
Split into format_date() and
get_due_date() helpers for modularity
\end{lstlisting} \\
\hline

2 &
\begin{lstlisting}
class SignUpSheetController < ApplicationController
  def create
    topic = Topic.new(params[:topic])
    topic.save
  end
end
\end{lstlisting} &
\begin{lstlisting}
# Controller simplified
topic = TopicService.new.create(topic_params)
\end{lstlisting} &
\begin{lstlisting}
class SignUpSheetController
  def create
    topic = TopicManager.new(params).create
  end
end
\end{lstlisting} &
\begin{lstlisting}
class SignUpSheetController
  def initialize(topic_service)
    @topic_service = topic_service
  end
  def create
    @topic_service.create(params)
  end
end
\end{lstlisting} \\
\hline
3 &
\begin{lstlisting}
class Assignment < ApplicationRecord
  include Scoring
  include ReviewAssignment
  include QuizAssignment
  include AssignmentHelper
end
\end{lstlisting} &
\begin{lstlisting}
# Refactored
class Assignment < ApplicationRecord
  include Assignable
  include Reviewable
end
\end{lstlisting} &
\begin{lstlisting}
class Assignment < ApplicationRecord
  def calculate_score
    ScoringService.new(self).calculate
  end
end
\end{lstlisting} &
\begin{lstlisting}
module AssignmentFeatures
  include Scoring
  include ReviewAssignment
end
class Assignment < ApplicationRecord
  include AssignmentFeatures
end
\end{lstlisting} \\
\hline
\end{tabular}
\label{tab:rag_comparison}
\end{table*}

The RAG approach was implemented by providing a curated knowledge base of core design principles as context to enhance LLM prompts adapted primarily from instructional material available at \cite{saasbook}. This knowledge base included detailed descriptions of SOLID principles, DRY, KISS, YAGNI, PoLA, Law of Demeter, Separation of Concerns\cite{Hursch1995}, and common design patterns. This contextual information was provided to the LLMs during the prompting process to ground their analysis in established software engineering principles.

\subsection{RAG Results Analysis}
The integration of RAG significantly improved the quality and specificity of feedback across all three models. Table VII illustrates a comparative breakdown of refactoring examples suggested by GPT-4 (RAG), Claude, and DeepSeek for the same code snippets. Each model approached the improvements differently:

% The integration of RAG significantly improved the quality and specificity of feedback across all models. Key improvements observed include:

% \begin{itemize}
%     \item \textit{Enhanced context-awareness:} Models demonstrated a stronger ability to connect code issues to specific design principles when provided with the RAG knowledge base.
    
%     \item \textit{Principle-specific references:} LLMs provided direct references to design principles, creating educational opportunities within the feedback itself.
    
%     \item \textit{More actionable refactoring suggestions:} Models offered concrete code examples for implementing improvements.
% \end{itemize}

% Comparative analysis of model-specific RAG responses:

\emph{GPT-4 with RAG} focused on parameterizing methods, refactoring to use keyword arguments, and restructuring helper functions (e.g., splitting logic in \lstinline{check_topic_due_date_val}).

% GPT-4's analysis with RAG focused on practical implementation of theoretical principles. It excelled at identifying Liskov Substitution Principle violations, providing concrete examples such as improper inheritance hierarchies.

% GPT-4 also demonstrated strength in recommending practical refactorings aligned with the Single Responsibility Principle, as shown in this example for a helper method:

% \lstset{
%   basicstyle=\ttfamily\small,
%   breaklines=true,
%   frame=none,
%   columns=fullflexible,
%   keepspaces=true,
%   showstringspaces=false,
%   captionpos=b
% }

% \begin{lstlisting}[frame=none, caption={Refactoring for SRP}, basicstyle=\footnotesize, captionpos=b, float=htb]

% //Before Refactoring:
% Method check_due_date:
%     Retrieve due date
%     Format due date
%     Return formatted date

% //After Refactoring:
% Method get_due_date:
%     Retrieve due date
%     Return due date

% Method format_date(due_date):
%     Format due date
%     Return formatted date

% Method check_due_date:
%     Call get_due_date
%     Call format_date with due date
%     Return formatted date

% \end{lstlisting}

\textit{Claude with RAG} emphasized architectural and responsibility separation by moving business logic into dedicated service classes (e.g., \lstinline{TopicDeadlineService} or \lstinline{TopicManager}).
% Claude's RAG-enhanced analysis emphasized architectural concerns and demonstrated a thorough understanding of responsibility distribution across classes. Its suggestions included moving business logic out of controllers and into well-defined object classes. For instance, refactoring topic-related functionality into a dedicated Topic class allowed for clearer separation of concerns, reflecting a robust application of the Single Responsibility Principle:

% \begin{lstlisting}[caption={SignUpSheetController Class}, frame=none, basicstyle=\footnotesize, captionpos=b, float=htb]
% class SignUpSheetController < ApplicationController
%   def create_topic
%     topic = Topic.new(topic_params)
%     if topic.save
%       // success flow
%     else
%       // error handling
%     end
%   end
% end
% \end{lstlisting}

% \begin{lstlisting}[caption={Topic Class}, frame=none, basicstyle=\footnotesize, captionpos=b, float=htb]
% Class Topic:
%     Attributes: title, max_choosers

%     Constructor(params):
%         title = params.title
%         max_choosers = params.max_choosers

%     Method save():
%         Persist topic

%     Method valid():
%         Validate topic attributes

% \end{lstlisting}

% Claude's recommendations were particularly valuable for high-level design decisions. By relocating logic into domain-specific object classes like Topic, the codebase becomes more modular, easier to test, and aligned with object-oriented best practices.

\textit{DeepSeek with RAG} delivered detailed code-level restructuring, recommending dependency injection patterns and modularizing concerns (e.g., grouping \lstinline{Scoring} and \lstinline{ReviewAssignment} modules into \lstinline{AssignmentFeatures}).

Table \ref{tab:rag_comparison} also shows that while GPT-4 suggested small, targeted improvements for readability and maintainability, Claude provided higher-level design abstractions, and DeepSeek combined framework-aware optimizations with systematic modularization.

\subsection{Impact on Feedback Quality}

% The incorporation of RAG substantially improved feedback quality across several dimensions:

% \begin{enumerate}
% \item\textit{Principle-aligned feedback:} RAG-enhanced prompts improved precision and consistency by grounding responses in established design principles.

% \item \textit{Educational explanations:} Models provided clearer reasoning behind suggestions, creating better learning opportunities for students.

% \item\textit{Domain-specific insights:} Models demonstrated improved framework-aware feedback (e.g., Ruby on Rails), with actionable examples and refactorings\cite{Hartl2022,FowlerRubyEdition}.

% \end{enumerate}

% The code examples provided by the RAG-enhanced models demonstrate a clear understanding of both general design principles and framework-specific best practices. This suggests that RAG is particularly valuable in educational contexts where students need concrete examples to understand abstract design concepts.

The comparison in Table \ref{tab:rag_comparison} highlights how each LLM excels in distinct dimensions of code quality:
\begin{enumerate}

\item Principle-aligned feedback:
GPT-4 was most effective at tying its suggestions to principles like Single Responsibility and DRY by directly improving code structure, while Claude leveraged Separation of Concerns by shifting logic out of controllers. DeepSeek uniquely introduced dependency injection to decouple components, demonstrating a stronger focus on low coupling and testability.
\item Educational and actionable explanations:
Claude’s suggestions (e.g., encapsulating logic in TopicManager) provided high-level architectural reasoning, helping developers understand why a refactor improves maintainability. GPT-4, in contrast, provided concise yet actionable refactor snippets (e.g., converting positional to keyword arguments). DeepSeek excelled in showing step-by-step code reorganization that maps directly to Rails best practices.
\item Diversity of design insights:
Table \ref{tab:rag_comparison} shows that all three models identified distinct sets of improvements for the same snippet, demonstrating the value of combining multi-model outputs for a more comprehensive review pipeline.

In summary, Table \ref{tab:rag_comparison} acts as both a diagnostic and comparative tool, capturing each model’s unique perspective on enforcing design principles. This comparative approach improves coverage of design feedback, ensuring both micro-level code improvements and macro-level architectural refinements are addressed.

\end{enumerate}

\section{Implementation Recommendations}
To maximize the effectiveness of LLM-assisted code feedback and evaluation, the following implementation strategies are recommended:

\subsection{Optimized prompt engineering} Crafting high-quality prompts is critical for eliciting useful responses. Effective prompts should: \begin{itemize} \item Clearly define the evaluation criteria, such as readability, modularity, maintainability, and adherence to specific design principles (e.g., SOLID, DRY). \item Combine targeted and holistic scopes, enabling the model to focus on specific functions while also assessing overall architectural decisions. \item Request structured responses, such as bullet points or categorized feedback, which improve clarity and make model outputs easier to compare and act upon. \item Use iterative prompting, allowing refinement of responses through follow-up queries or clarification prompts. \end{itemize}

\subsection{Leveraging multi-model feedback} Different LLMs excel in various areas. A multi-model approach can: \begin{itemize} \item Provide diverse perspectives on code quality and design. \item Allow comparison of model reasoning styles, particularly useful in educational settings. \item Enable cross-validation of feedback for higher confidence in model suggestions. \end{itemize}

\subsection{Incorporating RAG for contextual relevance} Retrieval-Augmented Generation (RAG) enhances model performance by grounding outputs in relevant external knowledge. Key recommendations include: \begin{itemize} \item Building a curated knowledge base of course materials, best practices, and past code reviews. \item Balancing retrieval with generation, ensuring responses remain adaptive but well-informed. \item Contextual injection of retrieved information, guiding models toward domain-accurate and pedagogically aligned feedback\cite{ChenReadingWikipedia}. \end{itemize}

\section{Conclusion and Future Work}
\subsection{Conclusion} LLMs offer a transformative opportunity to scale and personalize feedback in programming education and software engineering practices. Through strategic prompt design, careful model selection, and intelligent integration of retrieval mechanisms, these models can provide detailed, criterion-aligned evaluations that enhance learning outcomes and developer productivity. However, effective deployment requires thoughtful implementation—balancing model capabilities, prompt structure, and domain context.

\subsection{Future Work} To further improve the impact and reliability of LLM-powered systems, future efforts should focus on: \begin{itemize} \item Expanding the knowledge base used in RAG systems with more granular tagging and contextual metadata. \item Embedding LLMs directly within development tools to enable real-time, inline feedback as users code. \item Improving model accuracy and grounding, especially in domain-specific or educational contexts, to reduce hallucination and increase trust. \item Personalizing feedback based on user profiles, learning goals, and historical performance for adaptive support. \item Evaluating learning outcomes in longitudinal studies to assess how LLM feedback influences student understanding and software quality over time. \end{itemize}

\bibliographystyle{IEEEtran}
\bibliography{references}    % your .bib file

\end{document}